\documentstyle[aps]{revtex}

\begin{document}
\author{Edward Gerjuoy}
\address{Dept. of Physics, University of Pittsburgh, Pittsburgh, PA 15260}
\title{Lower Bound on Entanglement of Formation for the Qubit-Qudit System }
\date{1/3/03}
\maketitle

\begin{abstract}
Wootters [PRL $80$, 2245 (1998)] has derived a closed formula for the
entanglement of formation (EOF) of an arbitrary mixed state in a system of
two qubits. There is no known closed form expression for the EOF of an
arbitrary mixed state in any system more complicated than two qubits. This
paper, via a relatively straightforward generalization of Wootters' original
derivation, obtains a closed form lower bound on the EOF of an arbitary
mixed state of a system composed of a qubit and a qudit (a {\it d}-level
quantum system, with {\it d} $\geq $ 3). The derivation of the lower bound
is detailed for a system composed of a qubit and a qutrit ({\it d} = 3); the
generalization to {\it d} $>$ 3 then follows readily.

PACS number(s): $03.65.$Ud, 03.67.-a
\end{abstract}

\section{Introduction.}

As Wootters [1] discusses, the ''entanglement of formation'' (defined below,
hereinafter simply ''EOF'') is one of the more commonly employed measures of
the entanglement of an arbitrary mixed state of a bipartite quantum system,
i.e., a quantum system composed of two and only two quantum subsystems; for
pure states of bipartite quantum systems, the von Neumann entropy (also
defined below) is the standard measure of entanglement. Indeed Wootters [2],
taking advantage of the fact that the von Neumann entropy of a pure state of
two qubits is a function of a single real parameter only, has derived a
closed formula for the EOF of an arbitrary mixed state in a system of two
qubits. There is no known closed form expression for the EOF of an arbitrary
mixed state in any bipartite system more complicated than two qubits. As
elaborated below, however, the von Neumann entropy of a pure state composed
of a qudit (any wavefunction is a linear combination of at most {\it d}
orthonormal eigenfunctions) and a qubit (a qudit in the special case{\it \ d
= }2) also is a function of a single real parameter only.

\smallskip This paper therefore has sought to generalize Wootters' two qubit
derivation [2], so as to obtain a closed formula for the EOF of an arbitrary
mixed state of the qubit-qudit system. This attempt has not been successful,
for reasons that will be manifest, but it has proved possible to derive a
closed form lower bound on the EOF of an arbitrary qubit-qudit mixed state.
Presenting this derivation is the primary objective of this paper. The
derivation will be detailed for an arbitrary mixed state of a qubit and a
qutrit (a qudit in the special case {\it d} = 3); the generalization to the
qubit-qudit (with {\it d} 
\mbox{$>$}%
3) case then follows readily.

\section{The Qubit-Qutrit System.}

Let $\rho $ denote an arbitrary density matrix of a bipartite quantum system
S composed of a qubit and a qutrit. Unless otherwise stated, it is assumed
throughout this paper that $\rho ^{2}$ $\neq $ $\rho ,$ i.e., that $\rho $
does not fortuitously happen to represent a pure state. Then the mixed state
density matrix $\rho $ can be written in the form 
\begin{equation}
\rho =\sum_{\alpha =1}^{N}p_{\alpha }\Psi _{\alpha }\Psi _{\alpha }^{\dagger
}
\end{equation}
(i.e., can be ''decomposed'') in an infinite number of ways [3]. In Eq. (1): 
$\rho $ is a 6$\times 6$ matrix; each $p_{\alpha }$ is a real number $>$ 0; $%
\Sigma _{\alpha }p_{\alpha }=$ 1; 2 $\leq $ {\it r }$\leq 6$ is the rank of $%
\rho ,$ i.e., the number of non-zero eigenvalues of $\rho ;$ {\it N} is $%
\geq r$ and can be an arbitrarily large integer; the dagger denotes the
adjoint; Dirac notation, which in this writer's opinion beclouds the matrix
structure of the pertinent mathematics, deliberately has been avoided (as it
will be during the remainder of this paper); and each $\Psi _{\alpha }$ is a
normalized system S wave function, expressible as

\begin{equation}
\Psi _{\alpha }=\sum_{i=1}^{2}\sum_{j=1}^{3}a_{ij}^{\alpha }u_{i}v_{j},
\end{equation}
where the $u_{i}$ and $v_{j}$ are orthonormal eigenfunctions of the qubit
and qutrit respectively. In general the different $\Psi _{\alpha }$ forming
any given decomposition (1) are not orthogonal to each other, but for each $%
\Psi _{\alpha }$ the coefficients $a_{ij}^{\alpha }$ satisfy 
\begin{equation}
\sum_{i,j}\left| a_{ij}^{\alpha }\right| ^{2}=\Psi _{\alpha }^{\dagger }\Psi
_{\alpha }=1.
\end{equation}

According to the Schmidt decomposition theorem [4], for any $\Psi _{\alpha }$
Eq. (2) can be replaced by 
\begin{equation}
\Psi _{\alpha }=\sum_{k}c_{k}^{\alpha }\overline{u}_{k}^{\alpha }\overline{v}
_{k}^{\alpha },
\end{equation}
where: the $\overline{u}_{k}^{\alpha }$ are orthonormal linear combinations
of the $u_{i};$ the $\overline{v}_{k}^{\alpha }$ are orthonormal linear
combinations of the $v_{j};$ and the quantities $c_{k}^{\alpha }$ are
non-negative real numbers. As the notation indicates, all the $\overline{u}
_{k}^{\alpha },$ $\overline{v}_{k}^{\alpha },$ and $c_{k}^{\alpha }$ are $%
\alpha $-dependent. Since there are at most two orthonormal $\overline{u}%
_{k}^{\alpha }$ for any $\alpha $, there are at most two non-vanishing $%
c_{k}^{\alpha }$ in the Schmidt expansion (4) for $\Psi _{\alpha },$ even
though there can be three orthonormal $\overline{v}_{k}^{\alpha }.$
Correspondingly the normalization relation (3) reduces to 
\begin{equation}
\sum_{k=1}^{2}(c_{k}^{\alpha })^{2}=1,
\end{equation}
wherein the range of each $c_{k}^{\alpha }$ is 0 $\leq $ $c_{k}^{\alpha
}\leq 1,$ of course.

The von Neumann entropy associated with the qubit or qutrit subsystem
reduced density matrices that can be constructed from the wave function $%
\Psi _{\alpha }$ of Eq. (2) is [1] 
\begin{equation}
E(\Psi _{\alpha })=-\sum_{k=1}^{2}(c_{k}^{\alpha })^{2}\log
_{2}(c_{k}^{\alpha })^{2},
\end{equation}
where the $c_{k}^{\alpha }$ are the coefficients in the expansion (4).
Because of Eq. (5) $E(\Psi _{\alpha })$ actually is a function of a single
parameter\thinspace only, which conveniently can be chosen to be{\rm \ } 
\begin{equation}
C_{\alpha }=2c_{1}^{\alpha }c_{2}^{\alpha };
\end{equation}
evidently 0 $\leq $ $C_{\alpha }\leq $ 1. Eq. (5) then yields, for use in
Eq. (6), 
\begin{eqnarray}
(c_{1}^{\alpha })^{2} &=&\frac{1}{2}(1+\sqrt{1-C_{\alpha }^{2}}) \\
(c_{2}^{\alpha })^{2} &=&\frac{1}{2}(1-\sqrt{1-C_{\alpha }^{2}})
\end{eqnarray}
where we now are choosing $c_{1}^{\alpha }\geq c_{2}^{\alpha }.$ Therewith
we can rewrite Eq. (6) as 
\begin{equation}
E(\Psi _{\alpha })={\Large \varepsilon }(C_{\alpha }),
\end{equation}
with $\varepsilon (C_{\alpha })$ the expression obtained after substitution
of Eqs. (8)-(9) into the right side of Eq. (6).

The EOF of $\rho $ is defined to be [1] 
\begin{equation}
{\bf E(}\rho {\bf )=}\text{ min}\sum_{\alpha =1}^{N}p_{\alpha }E(\Psi
_{\alpha })\equiv \text{min}\sum_{\alpha =1}^{N}p_{\alpha }{\Large %
\varepsilon }(C_{\alpha }),
\end{equation}
minimized over all possible decompositions of $\rho $ of the form (1).%
\footnote{%
We shall not concern ourselves here with the difference between the
minimization operation indicated in Eq. (11) and the infimum operation
favored in Wootters' review [1].} Also define the similarly minimized
quantity 
\begin{equation}
{\bf C(}\rho {\bf )=}\text{ min}\sum_{\alpha =1}^{N}p_{\alpha }C_{\alpha }.
\end{equation}
Because [1] $\varepsilon (C_{\alpha })$ is a monotonically increasing convex
function of $C_{\alpha }$ in its range 0 $\leq $ $C_{\alpha }\leq $ 1, 
\begin{equation}
{\Large \varepsilon }{\large (}\sum_{\alpha }p_{\alpha }C_{\alpha }{\large )}%
\leq \sum_{\alpha }p_{\alpha }{\Large \varepsilon }(C_{\alpha }),
\end{equation}
implying [1] 
\begin{equation}
{\Large \varepsilon }[{\bf C}(\rho )]\leq {\bf E(}\rho {\bf ).}
\end{equation}

In order to obtain a closed formula for ${\bf E(}\rho {\bf )}$ from Eq.
(14), or even merely a bound on {\bf E}($\rho ),$ it is necessary to
establish an explicit connection between $C_{\alpha }$ and $\Psi _{\alpha },$
well beyond the implicit connection inherent in Eq. (4). Wootters' ability
to establish such a connection, via introduction of the ''concurrence'' [2],
is a key feature of his successful derivation of a closed formula for the
EOF of an arbitrary mixed state in the two qubit case. At this juncture,
therefore, it is helpful to briefly review Wootters' definition and
application of the concurrence.

\subsection{The Concurrence. Wootters' Two Qubit Derivation.}

Until further notice we are considering the two qubit system only, rather
than the qubit-qutrit system S to which the preceding Eqs. (1)-(14) pertain. 
$\rho $ now is a 4$\times 4$ matrix, not 6$\times 6;$ the range of {\it r }%
in Eq. (1) now is 2 $\leq r\leq 4;$ and the subscript {\it j }in Eqs. (2)
and (3) now is permitted to range over the values 1,2 only, rather than the
values 1,2,3. With these understandings Eqs. (1)-(14) are applicable to the
two qubit system without further alteration. Then in effect the concurrence
of any two qubit wave function $\Psi _{\alpha }$ is defined as [2] 
\begin{equation}
Con(\Psi _{\alpha })=\left| \Psi _{\alpha }^{\dagger }{\bf S}\Psi _{\alpha
}^{*}\right| ,
\end{equation}
wherein: $\Psi _{\alpha }$ is a column matrix whose components $\Psi
_{m\alpha },$ $m=1,2,3,4,$ are respectively the coefficients $a_{11}^{\alpha
},a_{12}^{\alpha },a_{21}^{\alpha },a_{22}^{\alpha }$ appearing in the two
qubit version of Eq. (2); the asterisk denotes the complex conjugate; the
notation is consistent with Eq. (1), with $\Psi _{\alpha }^{\dagger }$ the
row matrix that is the transpose of $\Psi _{\alpha }^{*};$ and {\bf S} is
the matrix 
\begin{equation}
{\bf S}=\left( 
\begin{array}{cccc}
0 & 0 & 0 & -1 \\ 
0 & 0 & 1 & 0 \\ 
0 & 1 & 0 & 0 \\ 
-1 & 0 & 0 & 0
\end{array}
\right) .
\end{equation}
Evidently 
\begin{equation}
Con(\Psi _{\alpha })=\left| 2(a_{11}^{\alpha }a_{22}^{\alpha
}-a_{21}^{\alpha }a_{12}^{\alpha })\right| .
\end{equation}

The numerical coefficients $a_{ij}^{\alpha }$ in Eq. (2) now form a square
matrix 
\begin{equation}
A_{\alpha }=\left( 
\begin{array}{ll}
a_{11}^{\alpha } & a_{12}^{\alpha } \\ 
a_{21}^{\alpha } & a_{22}^{\alpha }
\end{array}
\right) ,
\end{equation}
to which the singular value decomposition theorem [5,6] 
\begin{equation}
A_{\alpha }=U_{\alpha }\Delta _{\alpha }V_{\alpha }
\end{equation}
applies. In Eq. (19): $U_{\alpha }$ and $V_{\alpha }$ are ($\alpha -$%
dependent) unitary matrices; the diagonal matrix 
\begin{equation}
\Delta _{\alpha }=\left( 
\begin{array}{ll}
c_{1}^{\alpha } & 0 \\ 
0 & c_{2}^{\alpha }
\end{array}
\right) ;
\end{equation}
and $c_{1}^{\alpha },c_{2}^{\alpha }$ are the coefficients appearing in Eq.
(4) with (as previously specified) $c_{1}^{\alpha }\geq c_{2}^{\alpha }.$
Eqs. (18)-(20) immediately imply the determinantal equality $\left| \det
(A_{\alpha })\right| =\left| \det (\Delta _{\alpha })\right| ,$ i.e., 
\begin{equation}
\left| a_{11}^{\alpha }a_{22}^{\alpha }-a_{21}^{\alpha }a_{12}^{\alpha
}\right| =c_{1}^{\alpha }c_{2}^{\alpha }.
\end{equation}
It follows that $Con(\Psi _{\alpha })$ is precisely equal to the parameter $%
C_{\alpha }$ defined in Eq. (7). Thus the two qubit version of Eq. (12)
becomes 
\begin{equation}
{\bf C(}\rho {\bf )=\min }\sum_{\alpha =1}^{N}p_{\alpha }\left| \Psi
_{\alpha }^{\dagger }{\bf S}\Psi _{\alpha }^{*}\right| \equiv {\bf \ \min }
\sum_{\alpha =1}^{N}p_{\alpha }\left| \tilde{\Psi}_{\alpha }{\bf S}\Psi
_{\alpha }\right| ,
\end{equation}
where $\tilde{\Psi}_{\alpha }$ is the transpose of $\Psi _{\alpha }.$
Evidently ${\bf C(}\rho {\bf )}$ is the minimum average concurrence that any
decomposition of $\rho $ can attain.

Starting from Eq. (22), Wootters [2] obtains his closed form expression for $%
{\bf C(}\rho {\bf )}$ as follows (in effect). Eq. (1) shows the matrix $\rho 
$ is not only Hermitian but also is positive semidefinite, meaning [10] that 
$\Theta ^{\dagger }\rho \Theta \geq 0$ for any two qubit wave function $%
\Theta .$ Thus $\rho $ can be brought to diagonal form $\Delta _{\rho }$ by
a unitary transformation {\it U}$_{\rho },$%
\begin{equation}
\rho =U_{\rho }\Delta _{\rho }U_{\rho }^{\dagger }=\sum_{s=1}^{4}\mu
_{s}w_{s}w_{s}^{\dagger },
\end{equation}
wherein the $\mu _{s}$ ($\mu _{1}\geq \mu _{2}\geq \mu _{3}\geq \mu _{4}\geq
0)$ are the eigenvalues of $\rho $ and the $w_{s}$ (components $w_{ms},$ 
{\it m} = 1,2,3,4) comprising the columns of $U_{\rho }$ are a corresponding
set of orthonormal eigenfunctions. Because $\rho $ is a density matrix,
i.e., has trace {\it Tr(}$\rho )=1,$ Eq. (23) has the form of Eq. (1), i.e.,
Eq. (23) provides a special decomposition of $\rho .$ Introducing {\it W}$%
_{s}=\sqrt{\mu _{s}}w_{s}$, and defining {\bf W }to be the 4$\times 4$
square matrix whose elements are {\it W}$_{ms},$ Eq. (23) takes the form 
\begin{equation}
\rho =\sum_{s=1}^{4}W_{s}W_{s}^{\dagger }\equiv {\bf WW}^{\dagger }
\end{equation}
wholly consistent with matrix notation. Similarly, introducing $\Phi
_{\alpha }=\sqrt{p_{\alpha }}\Psi _{\alpha }$ and defining ${\bf \Phi }$ to
be the 4$\times N$ rectangular matrix (4 rows, {\it N} columns) whose
elements are $\Phi _{m\alpha },$ Eq. (1) becomes 
\begin{equation}
\rho =\sum_{\alpha =1}^{N}\Phi _{\alpha }\Phi _{\alpha }^{\dagger }\equiv 
{\bf \Phi \Phi }^{\dagger },
\end{equation}
again wholly consistent with matrix notation. Furthermore, starting with any
set of column matrices $\Phi _{\alpha }$ satisfying Eq. (25), defining $\Psi
_{\alpha }=\Phi _{\alpha }/\sqrt{(\Phi _{\alpha }^{\dagger }\Phi _{\alpha })}%
,$ and remembering {\it Tr(}$\rho )=1,$ it can be seen that such a set $\Psi
_{\alpha }$ satisfies Eq. (1). In other words not only does any
decomposition of $\rho $ provide a set $\Phi _{\alpha }$ satisfying Eq.
(25), but also any set $\Phi _{\alpha }$ satisfying Eq. (25) provides a
decomposition of $\rho .$

Next let {\bf R} be any 4$\times N$ matrix satisfying 
\begin{equation}
{\bf RR}^{\dagger }={\bf I,}
\end{equation}
where ${\bf I}$ is the unit matrix. Eq. (26) implies that the 4 rows of {\bf %
R }(but not necessarily its {\it N} columns unless {\bf R} happens to be
square) form a set of {\it N}-component orthonormal vectors. Then, as
Wootters [2] observes, the matrix ${\bf \Phi }$ defined by 
\begin{equation}
{\bf \Phi }={\bf WR}
\end{equation}
satisfies Eq. (25). Moreover it can be proved [3] that every ${\bf \Phi }$
satisfying Eq. (25) necessarily satisfies Eq. (27) for some {\bf R} obeying
Eq. (26). It follows that Eq. (22) can be replaced by 
\begin{equation}
{\bf C(}\rho {\bf )=\min }\sum_{\alpha =1}^{N}\left| ({\bf \tilde{\Phi}S\Phi 
})_{\alpha \alpha }\right| ={\bf \min }\sum_{\alpha =1}^{N}\left| ({\bf 
\tilde{R}\tilde{W}SWR})_{\alpha \alpha }\right| ,
\end{equation}
minimized over all possible {\bf R} satisfying Eq. (26). Because${\bf \ S}$
defined by Eq. (18) is symmetric, the square matrix ${\bf \tilde{W}SW}$
appearing in Eq. (28) also is symmetric. Consequently, via an extension of
the singular value decomposition theorem to symmetric matrices [9], 
\begin{equation}
{\bf \tilde{W}SW=}\tilde{U}_{W}\Delta _{W}U_{W},
\end{equation}
where $U_{W}$ is unitary and $\Delta _{W}$ is a diagonal matrix whose
diagonal elements 
\begin{equation}
\lambda _{1}\geq \lambda _{2}\geq \lambda _{3}\geq \lambda _{4}\geq 0
\end{equation}
are the positive square roots of the assuredly non-negative real eigenvalues
of the Hermitian matrix 
\begin{equation}
{\bf \tilde{W}SW(\tilde{W}SW)}^{\dagger }={\bf \tilde{W}SWW}^{\dagger }{\bf %
SW}^{*}={\bf \tilde{W}S}\rho {\bf SW}^{*}.
\end{equation}
It can be seen that the eigenvalues of ${\bf \tilde{W}S}\rho {\bf SW^{*}}$
coincide with the eigenvalues of ${\bf W}^{*}{\bf \tilde{W}S}\rho {\bf S=}%
\rho {\bf ^{*}S}\rho {\bf S,}$ as well as with the eigenvalues of $\rho {\bf %
S}\rho ^{*}{\bf S}$ and $\rho ^{1/2}{\bf S}\rho ^{*}{\bf S}\rho ^{1/2};$ the
matrix $\rho ^{1/2}{\bf S}\rho ^{*}{\bf S}\rho ^{1/2},$ like ${\bf \tilde{W}S%
}\rho {\bf SW^{*}},$ is obviously Hermitian and positive semidefinite.

Substituting Eq. (29) into Eq. (28) fixes our attention on the matrix {\bf Q 
}={\bf \ }$U_{W}{\bf R,}$ which like {\bf R }is a 4$\times N$ rectangular
matrix. Also since 
\begin{equation}
{\bf QQ}^{\dagger }=U_{W}{\bf RR}^{\dagger }U_{W}^{\dagger
}=U_{W}U_{W}^{\dagger }={\bf I,}
\end{equation}
the four rows of {\bf Q}, again like the four rows of {\bf R, }form a set of
N-component orthonormal vectors. Moreover to every {\bf Q} satisfying Eq.
(32) there corresponds an {\bf R} = $U_{W}^{\dagger }{\bf Q}$ satisfying Eq.
(26). Thus Eq. (28) simplifies to 
\begin{equation}
{\bf C(}\rho {\bf )=\min }\sum_{\alpha =1}^{N}\left| ({\bf \tilde{Q}}\Delta
_{W}{\bf Q})_{\alpha \alpha }\right| ={\bf \min }\sum_{\alpha =1}^{N}\left|
\sum_{m=1}^{4}\lambda _{m}{\bf Q}_{m\alpha }^{2}\right| ,
\end{equation}
minimized over all possible ${\bf Q}$ satisfying Eq. (32). In fact (as
Wootters [2] observes), because of the absolute value signs the minimization
in Eq. (33) can be restricted without loss of generality to matrices {\bf Q }
which not only satisfy Eq. (32) but for which also every {\bf Q}$_{1\alpha }$
is real and $\geq 0.$

The trace of the matrix ${\bf \tilde{Q}}\Delta _{W}{\bf Q}$ obeys 
\begin{equation}
\left| \text{Tr}({\bf \tilde{Q}}\Delta _{W}{\bf Q)}\right| =\left|
\sum_{\alpha =1}^{N}\sum_{m=1}^{4}\lambda _{m}{\bf Q}_{m\alpha }^{2}\right|
=\left| \sum_{m=1}^{4}\lambda _{m}\left( \sum_{\alpha =1}^{N}{\bf Q}%
_{m\alpha }^{2}\right) \right| \leq \sum_{\alpha =1}^{N}\left|
\sum_{m=1}^{4}\lambda _{m}{\bf Q}_{m\alpha }^{2}\right| .
\end{equation}
Therefore, now restricting ${\bf Q}$ as described at the end of the
preceding paragraph, Eqs. (33) and (34) imply 
\begin{equation}
{\bf \min }\left| \lambda _{1}+\lambda _{2}\sum_{\alpha =1}^{N}{\bf Q}%
_{2\alpha }^{2}+\lambda _{3}\sum_{\alpha =1}^{N}{\bf Q}_{3\alpha
}^{2}+\lambda _{4}\sum_{\alpha =1}^{N}{\bf Q}_{4\alpha }^{2}\right| \leq 
{\bf C(}\rho ),
\end{equation}
wherein the coefficients of $\lambda _{2},\lambda _{3},\lambda _{4}$ are
complex numbers having absolute values $\leq 1.$ It follows that if 
\begin{equation}
\lambda _{1}-\lambda _{2}-\lambda _{3}-\lambda _{4}\geq 0,
\end{equation}
then the smallest possible value of the minimum on the left side of Eq.
(35), which minimum furnishes a lower bound to ${\bf C(}\rho ),$ is $\lambda
_{1}-\lambda _{2}-\lambda _{3}-\lambda _{4}.$ In fact this minimum actually
is as small as $\lambda _{1}-\lambda _{2}-\lambda _{3}-\lambda _{4}, $
because with the so-restricted 
\begin{equation}
{\bf Q\equiv Q}_{>}=\frac{1}{2}\left( 
\begin{array}{cccc}
1 & 1 & 1 & 1 \\ 
i & i & -i & -i \\ 
i & -i & i & -i \\ 
i & -i & -i & i
\end{array}
\right)
\end{equation}
the sums $\sum_{\alpha =1}^{N}{\bf Q}_{m\alpha }^{2}$ in Eq. (35) attain the
value -1 for {\it m} equal to each of 2, 3 and 4; with this same ${\bf \
Q\equiv Q}_{>},$ moreover, the right side of Eq. (34), which by virtue of
Eq. (33) furnishes an upper bound to ${\bf C(}\rho ),$ also equals $\lambda
_{1}-\lambda _{2}-\lambda _{3}-\lambda _{4}.$ Thus when Eq. (36) holds, Eqs.
(33) - (35) imply 
\begin{equation}
{\bf C(}\rho )=\lambda _{1}-\lambda _{2}-\lambda _{3}-\lambda _{4}.
\end{equation}

On the other hand, when 
\begin{equation}
\lambda _{1}-\lambda _{2}-\lambda _{3}-\lambda _{4}<0,
\end{equation}
it readily can be seen (again as Wootters [2] observes) that there always
exist real angles $\theta _{2},\theta _{3},\theta _{4}$ for which 
\begin{equation}
\left| \lambda _{1}+\lambda _{2}e^{2i\theta _{2}}+\lambda _{3}e^{2i\theta
_{3}}+\lambda _{4}e^{2i\theta _{4}}\right| =0.
\end{equation}
Correspondingly when Eq. (39) holds 
\begin{equation}
{\bf C(}\rho )=0
\end{equation}
because, using Eq. (40) together with the matrix 
\begin{equation}
{\bf Q\equiv Q}_{<}=\frac{1}{2}\left( 
\begin{array}{cccc}
1 & 1 & 1 & 1 \\ 
e^{i\theta _{2}} & e^{i\theta _{2}} & -e^{i\theta _{2}} & -e^{i\theta _{2}}
\\ 
e^{i\theta _{3}} & -e^{i\theta _{3}} & e^{i\theta _{3}} & -e^{i\theta _{3}}
\\ 
e^{i\theta _{4}} & -e^{i\theta _{4}} & -e^{i\theta _{4}} & e^{i\theta _{4}}
\end{array}
\right) ,
\end{equation}
one sees that the right side of Eq. (34) {\it and} the left side of Eq. (35)
equals zero. The matrix ${\bf Q}_{<},$ which also is restricted as described
above, was given (in essence) by Wootters [2].

Wootters [2] has embodied Eqs. (38) and (41) in the single equation 
\begin{equation}
{\bf C(}\rho )={\bf \max (}0{\bf ,}\lambda _{1}-\lambda _{2}-\lambda
_{3}-\lambda _{4}).
\end{equation}
Note, as Wootters [2] recognizes, that the matrix ${\bf Q}_{>}$ is not the
only 4$\times 4$ (restricted as described above) unitary ${\bf Q}$ which,
after insertion into Eqs. (33) - (35), implies Eq. (38) when Eq. (36) holds.
In particular if the 4$\times 4$ matrix ${\bf P}$ is orthogonal, i.e., is
real and satisfies ${\bf P\tilde{P}}$ = {\bf I, }then ${\bf Q}$ = ${\bf Q}%
_{>}{\bf P}$ can serve, because (i) {\it Tr}$({\bf \tilde{P}\tilde{Q}}%
_{>}\Delta _{W}{\bf Q}_{>}{\bf P)}$ = {\it Tr}(${\bf \tilde{Q}}_{>}\Delta
_{W}{\bf Q}_{>}{\bf P\tilde{P})}$ = {\it Tr}(${\bf \tilde{Q}}_{>}\Delta _{W}%
{\bf Q}_{>}{\bf ),}$ and (ii) the expressions $\sum_{m=1}^{4}\lambda _{m}%
{\bf Q}_{m\alpha }^{2}$ on the right side of Eq. (34) remain real for all $%
\alpha ;$ it is additionally required only that each of these expressions be
positive, as they are for ${\bf Q}$ = ${\bf Q}_{>},$ a requirement which
therefore always is achievable for some range of ${\bf P}$ as ${\bf P}$ is
continuously varied from the unit matrix {\bf I. }There also is a range of
so-restricted unitary 4$\times 4$ matrices, all of the form of Eq. (42),
which imply Eq. (41) when Eq. (39) holds, because in this circumstance
(assuming $\lambda _{4}\neq 0)$ there evidently is a range of angles $\theta
_{2},\theta _{3},\theta _{4}$ consistent with Eq. (40). It is additionally
evident that the matrices ${\bf Q}$ yielding Eq. (43) need not be square,
i.e., that the number {\it N} of terms $\Psi _{\alpha }\Psi _{\alpha
}^{\dagger }$ in a decomposition of $\rho $ yielding the minimum possible
average concurrence can be greater than 4; in particular, the real matrix $%
{\bf P}$ satisfying ${\bf P\tilde{P}}$ = {\bf I }introduced earlier in this
paragraph can be 4$\times N,$ with $N$ arbitrarily large. {\bf \ }

Actually the foregoing review of Wootters derivation of Eq. (43) has
implicitly assumed that the rank {\it r} of $\rho $ is 4. If any $\mu _{s}$
in Eq. (23) happens to be zero, i.e., if {\it r} is less than 4, the
corresponding column {\it W}$_{s}$ of {\bf W} is identically zero. But Eq.
(27) makes the elements of ${\bf \Phi }$ completely independent of the rows
of {\bf R }corresponding to identically zero columns of {\bf W}. It follows
that when {\it r} equals 2 or 3 Eq. (27) can yield a ${\bf \Phi }$
satisfying Eq. (25), i.e., can yield a decomposition of $\rho ,$ even though 
{\bf R} does not fully satisfy Eq. (26). Indeed the only rows of {\bf R}, as
well as of {\bf Q }={\bf \ }$U_{W}{\bf R}$ (it can be seen), which surely
form a set of N-component orthonormal vectors when {\it r} equals 2 or 3 are
those rows which do not correspond to the identically zero columns of {\bf W}%
. It can be shown, however, as Wootters [2] remarks, that whenever a $\mu
_{s}$ = 0 the corresponding $\lambda _{s}=0$ in Eq. (35); thus the possibly
deviating elements of {\bf Q} appear in Eq. (33 - (35) only as coefficients
of these vanishing $\lambda _{s},$ and cannot affect the validity of the
inferences we have drawn from those equations. In particular these possible
deviations of {\bf Q }from strict compliance with Eq. (32) do not alter the
conclusions that: (i) no decomposition of $\rho $ can yield a smaller
average concurrence $\sum_{\alpha }p_{\alpha }C_{\alpha }$ than is given by
Eq. (43), and (ii) there always is at least one decomposition which, via
Eqs. (37) and (42) together with {\bf R} = $U_{W}^{\dagger }{\bf Q,}$
actually does yield an average concurrence equal to the right side of Eq.
(43).

\subsection{Derivation of the EOF Lower Bound}

In the qubit-qutrit system, to which we now return, the analog of the matrix
on the right side of Eq. (18) for $\Psi _{\alpha }$ of Eq. (2) is 
\begin{equation}
\left( 
\begin{array}{lll}
a_{11}^{\alpha } & a_{12}^{\alpha } & a_{13}^{\alpha } \\ 
a_{21}^{\alpha } & a_{22}^{\alpha } & a_{23}^{\alpha }
\end{array}
\right) ,
\end{equation}
which no longer is square. The singular value decomposition theorem is
applicable to non-square matrices [5,6], but for the purpose of computing
the determinants of matrices, as was done in deriving Eq. (21), it is
necessary [7] to work with square matrices. Therefore the qubit-qutrit
analog of Eq. (18) will be taken to be 
\begin{equation}
A_{\alpha }=\left( 
\begin{array}{lll}
a_{11}^{\alpha } & a_{12}^{\alpha } & a_{13}^{\alpha } \\ 
a_{21}^{\alpha } & a_{22}^{\alpha } & a_{23}^{\alpha } \\ 
0 & 0 & 0
\end{array}
\right) .
\end{equation}
In essence we are proceeding as if Eq. (2) pertains to a system of two
qutrits in which, however, the only wave functions $\Psi _{\alpha }$ of
interest have zero projection on the eigenfunction {\it u}$_{3}$ of the
first qutrit. Eq. (19) now remains valid for this qubit-qutrit $A_{\alpha },$
but Eq. (20) must be replaced by 
\begin{equation}
\Delta _{\alpha }=\left( 
\begin{array}{lll}
c_{1}^{\alpha } & 0 & 0 \\ 
0 & c_{2}^{\alpha } & 0 \\ 
0 & 0 & 0
\end{array}
\right) .
\end{equation}
Correspondingly, in the qubit-qutrit case the equation $\left| \det
(A_{\alpha })\right| =\left| \det (\Delta _{\alpha })\right| $ yields merely
0 = 0, hardly a useful analog of Eq. (21).

There is a useful analog of Eq. (21), however, which is obtained as follows.
From Eq. (45) 
\begin{equation}
A_{\alpha }A_{\alpha }^{\dagger }=\left( 
\begin{array}{lll}
b_{11}^{\alpha } & b_{12}^{\alpha } & 0 \\ 
b_{21}^{\alpha } & b_{22}^{\alpha } & 0 \\ 
0 & 0 & 0
\end{array}
\right) ,
\end{equation}
where 
\begin{equation}
b_{11}^{\alpha }=\left| a_{11}^{\alpha }\right| ^{2}+\left| a_{12}^{\alpha
}\right| ^{2}+\left| a_{13}^{\alpha }\right| ^{2}
\end{equation}
\begin{equation}
b_{12}^{\alpha }=a_{11}^{\alpha }a_{21}^{\alpha *}+a_{12}^{\alpha
}a_{22}^{\alpha *}+a_{13}^{\alpha }a_{23}^{\alpha *}
\end{equation}
\begin{equation}
b_{21}^{\alpha }=b_{12}^{\alpha *}
\end{equation}
\begin{equation}
b_{22}^{\alpha }=\left| a_{21}^{\alpha }\right| ^{2}+\left| a_{22}^{\alpha
}\right| ^{2}+\left| a_{23}^{\alpha }\right| ^{2}.
\end{equation}
From Eq. (19), furthermore, 
\begin{equation}
A_{\alpha }A_{\alpha }^{\dagger }=U_{\alpha }\Delta _{\alpha }^{2}U_{\alpha
}^{\dagger },
\end{equation}
which expresses the singular value decomposition theorem result [6] that the
diagonal elements of $\Delta _{\alpha }^{2}$ are the eigenvalues of the
Hermitian matrix $A_{\alpha }A_{\alpha }^{\dagger }.$ The corresponding
eigenvalue equation implied by Eq. (52) is 
\begin{equation}
A_{\alpha }A_{\alpha }^{\dagger }-\lambda {\bf I}=U_{\alpha }\Delta _{\alpha
}^{2}U_{\alpha }^{\dagger }-\lambda {\bf I=}U_{\alpha }(\Delta _{\alpha
}^{2}-\lambda {\bf I)}U_{\alpha }^{\dagger }.
\end{equation}
Eq$.$ (53) yields 
\begin{equation}
\det (A_{\alpha }A_{\alpha }^{\dagger }-\lambda {\bf I)=\det }(\Delta
_{\alpha }^{2}-\lambda {\bf I).}
\end{equation}
Equating the coefficients of $\lambda $ on the two sides of Eq. (53) we find 
\begin{equation}
b_{11}^{\alpha }b_{22}^{\alpha }-b_{12}^{\alpha }b_{21}^{\alpha
}=(c_{1}^{\alpha })^{2}(c_{2}^{\alpha })^{2},
\end{equation}
a result which also follows directly from the theory [8] of the
characteristic polynomial associated with $A_{\alpha }A_{\alpha }^{\dagger
}. $ With the aid of Eqs. (48)-(51) and some algebraic manipulations, Eq.
(55) can be rewritten in the form 
\begin{equation}
\left| a_{11}^{\alpha }a_{22}^{\alpha }-a_{12}^{\alpha }a_{21}^{\alpha
}\right| ^{2}+\left| a_{11}^{\alpha }a_{23}^{\alpha }-a_{13}^{\alpha
}a_{21}^{\alpha }\right| ^{2}+\left| a_{12}^{\alpha }a_{23}^{\alpha
}-a_{13}^{\alpha }a_{22}^{\alpha }\right| ^{2}=(c_{1}^{\alpha
})^{2}(c_{2}^{\alpha })^{2}.
\end{equation}

\smallskip Eq. (56) is the desired qubit-qutrit analog of Eq. (21). Next
introduce the three matrices 
\begin{equation}
{\bf S}_{x}=\left( 
\begin{array}{llllll}
0 & 0 & 0 & 0 & 1 & 0 \\ 
0 & 0 & 0 & -1 & 0 & 0 \\ 
0 & 0 & 0 & 0 & 0 & 0 \\ 
0 & -1 & 0 & 0 & 0 & 0 \\ 
1 & 0 & 0 & 0 & 0 & 0 \\ 
0 & 0 & 0 & 0 & 0 & 0
\end{array}
\right)
\end{equation}
\begin{equation}
{\bf S}_{y}=\left( 
\begin{array}{llllll}
0 & 0 & 0 & 0 & 0 & 1 \\ 
0 & 0 & 0 & 0 & 0 & 0 \\ 
0 & 0 & 0 & -1 & 0 & 0 \\ 
0 & 0 & -1 & 0 & 0 & 0 \\ 
0 & 0 & 0 & 0 & 0 & 0 \\ 
1 & 0 & 0 & 0 & 0 & 0
\end{array}
\right)
\end{equation}
\begin{equation}
{\bf S}_{z}=\left( 
\begin{array}{llllll}
0 & 0 & 0 & 0 & 0 & 0 \\ 
0 & 0 & 0 & 0 & 0 & 1 \\ 
0 & 0 & 0 & 0 & -1 & 0 \\ 
0 & 0 & 0 & 0 & 0 & 0 \\ 
0 & 0 & -1 & 0 & 0 & 0 \\ 
0 & 1 & 0 & 0 & 0 & 0
\end{array}
\right) ,
\end{equation}
which can be thought of as qubit-qutrit analogs of Eq. (16). In terms of
these matrices Eq. (56) becomes 
\begin{equation}
\left( \left| \tilde{\Psi}_{\alpha }{\bf S}_{x}\Psi _{\alpha }\right|
^{2}+\left| \tilde{\Psi}_{\alpha }{\bf S}_{y}\Psi _{\alpha }\right|
^{2}+\left| \tilde{\Psi}_{\alpha }{\bf S}_{z}\Psi _{\alpha }\right|
^{2}\right) ^{1/2}=2c_{1}^{\alpha }c_{2}^{\alpha },
\end{equation}
where $\Psi _{\alpha }$ now is the column matrix whose six components $\Psi
_{m\alpha },$ $m=1$ to 6, are respectively the coefficients $a_{11}^{\alpha
},a_{12}^{\alpha },a_{13}^{\alpha },a_{21}^{\alpha },a_{22}^{\alpha
},a_{23}^{\alpha }$ appearing in Eqs. (2) and (45). Evidently the left side
of Eq. (60), like the two qubit concurrence of $\Psi _{\alpha }$ defined in
Eq. (15), precisely equals the parameter {\it C}$_{\alpha }$ defined in Eq.
(7). Thus the left side of Eq. (60) can be regarded as the qubit-qutrit
analog of the two qubit concurrence, and legitimately can be denoted by {\it %
Con}($\Psi _{\alpha }).$ The corresponding analog of Eq. (22), expressing
the minimum average concurrence any decomposition of $\rho $ can attain, is 
\begin{equation}
{\bf C(}\rho {\bf )=\min }\sum_{\alpha =1}^{N}p_{\alpha }Con(\Psi _{\alpha
})\equiv {\bf \min }\sum_{\alpha =1}^{N}p_{\alpha }\left( \left| \tilde{\Psi}%
_{\alpha }{\bf S}_{x}\Psi _{\alpha }\right| ^{2}+\left| \tilde{\Psi}_{\alpha
}{\bf S}_{y}\Psi _{\alpha }\right| ^{2}+\left| \tilde{\Psi}_{\alpha }{\bf S}%
_{z}\Psi _{\alpha }\right| ^{2}\right) ^{1/2}.
\end{equation}
Especially in view of the square roots therein, the right side of Eq. (61)
is sufficiently more complicated than the right side of Eq. (22) that this
writer has been unable to generalize Wootters' derivation to obtain a closed
formula for the EOF of an arbitrary qubit-qutrit mixed state. A lower bound
on this EOF can be derived, however, via consideration of the following
exercise:

We seek the minimum value {\it F}$_{\min }$ of 
\begin{equation}
F=\sum_{\alpha =1}^{N}(x_{\alpha }^{2}+y_{\alpha }^{2}+z_{\alpha }^{2})^{1/2}
\end{equation}
subject to the constraints 
\begin{equation}
\sum_{\alpha =1}^{N}x_{\alpha }=X,
\end{equation}
\begin{equation}
\sum_{\alpha =1}^{N}y_{\alpha }=Y,
\end{equation}
\begin{equation}
\sum_{\alpha =1}^{N}z_{\alpha }=Z,
\end{equation}
wherein all the {\it x}$_{\alpha },y_{\alpha },z_{\alpha }$ are real
variables $\geq 0,$ and {\it X, Y, Z} are given fixed quantities. Then
routine employment of Lagrange multipliers to take account of the
constraints finds that the right side of Eq. (62) has the single extremum
(which can be shown to be a minimum) 
\begin{equation}
F_{\min }=(X^{2}+Y^{2}+Z^{2})^{1/2}.
\end{equation}
Eq. (66) implies 
\begin{equation}
(X^{2}+Y^{2}+Z^{2})^{1/2}\leq \sum_{\alpha =1}^{N}(x_{\alpha }^{2}+y_{\alpha
}^{2}+z_{\alpha }^{2})^{1/2}
\end{equation}
whenever {\it x}$_{\alpha },y_{\alpha },z_{\alpha }$ are constrained by Eqs.
(63) - (65).

We now observe that Eqs. (23) and (24) remain valid in the qubit-qutrit
system, except that{\bf \ }the sum over {\it s} in those equations now runs
from 1 to 6, not merely 1 to 4; correspondingly {\bf W} now is a 6$\times 6$
matrix. Similarly Eqs. (25) - (27) also remain fully valid, except that {\bf %
R} and ${\bf \Phi }$ now are 6$\times N$ matrices, remembering of course
that $\Phi _{\alpha }=\sqrt{p_{\alpha }}\Psi _{\alpha }.$ In terms of the $%
\Phi _{\alpha }$ Eq. (61) takes the rather more convenient form 
\begin{equation}
{\bf C(}\rho {\bf )=\min }\sum_{\alpha =1}^{N}\left( \left| \tilde{\Phi}%
_{\alpha }{\bf S}_{x}\Phi _{\alpha }\right| ^{2}+\left| \tilde{\Phi}_{\alpha
}{\bf S}_{y}\Phi _{\alpha }\right| ^{2}+\left| \tilde{\Phi}_{\alpha }{\bf S}%
_{z}\Phi _{\alpha }\right| ^{2}\right) ^{1/2}.
\end{equation}
Any given decomposition of $\rho ,$ i.e., any given set $\Phi _{\alpha }$ in
the qubit-qutrit analog of Eq. (25), determines the values of the quantities 
{\it X}$_{D},Y_{D},Z_{D}$ defined by 
\begin{equation}
X_{D}=\sum_{\alpha =1}^{N}\left| \tilde{\Phi}_{\alpha }{\bf S}_{x}\Phi
_{\alpha }\right| ,
\end{equation}
\begin{equation}
Y_{D}=\sum_{\alpha =1}^{N}\left| \tilde{\Phi}_{\alpha }{\bf S}_{y}\Phi
_{\alpha }\right| ,
\end{equation}
\begin{equation}
Z_{D}=\sum_{\alpha =1}^{N}\left| \tilde{\Phi}_{\alpha }{\bf S}_{z}\Phi
_{\alpha }\right| .
\end{equation}
Identifying $\left| \tilde{\Phi}_{\alpha }{\bf S}_{x}\Phi _{\alpha }\right|
, $ $\left| \tilde{\Phi}_{\alpha }{\bf S}_{y}\Phi _{\alpha }\right| ,$ $%
\left| \tilde{\Phi}_{\alpha }{\bf S}_{z}\Phi _{\alpha }\right| $ in Eqs.
(69) - (71) with {\it x}$_{\alpha },y_{\alpha },z_{\alpha }$ respectively in
Eqs. (63) - (65), it is evident that Eq. (67) implies 
\begin{equation}
(X_{D}^{2}+Y_{D}^{2}+Z_{D}^{2})^{1/2}\leq \sum_{\alpha =1}^{N}\left( \left| 
\tilde{\Phi}_{\alpha }{\bf S}_{x}\Phi _{\alpha }\right| ^{2}+\left| \tilde{%
\Phi}_{\alpha }{\bf S}_{y}\Phi _{\alpha }\right| ^{2}+\left| \tilde{\Phi}%
_{\alpha }{\bf S}_{z}\Phi _{\alpha }\right| ^{2}\right) ^{1/2},
\end{equation}
where X$_{D},Y_{D},Z_{D}$ are defined by Eqs. (69) - (71). Eqs. (68) and
(72) in turn imply 
\begin{equation}
\min (X_{D}^{2}+Y_{D}^{2}+Z_{D}^{2})^{1/2}\leq {\bf C(}\rho {\bf ).}
\end{equation}

Next define for this qubit-qutrit case, in analogy with the first equality
in the two qubit Eq. (28), 
\begin{equation}
{\bf C}_{x}(\rho )={\bf \min }X_{D}={\bf \min }\sum_{\alpha =1}^{N}\left| 
\tilde{\Phi}_{\alpha }{\bf S}_{x}\Phi _{\alpha }\right| ,
\end{equation}
\begin{equation}
{\bf C}_{y}(\rho )={\bf \min }Y_{D}={\bf \min }\sum_{\alpha =1}^{N}\left| 
\tilde{\Phi}_{\alpha }{\bf S}_{y}\Phi _{\alpha }\right| ,
\end{equation}
\begin{equation}
{\bf C}_{z}(\rho )={\bf \min }Z_{D}={\bf \min }\sum_{\alpha =1}^{N}\left| 
\tilde{\Phi}_{\alpha }{\bf S}_{z}\Phi _{\alpha }\right| .
\end{equation}
Then Eq. (73) yields 
\begin{equation}
({\bf C}_{x}^{2}+{\bf C}_{y}^{2}+{\bf C}_{z}^{2})^{1/2}\leq {\bf C(}\rho 
{\bf ).}
\end{equation}
Furthermore, just as in Eq. (28), we now have 
\begin{equation}
{\bf C}_{x}{\bf (}\rho {\bf )=\min }\sum_{\alpha =1}^{N}\left| ({\bf \tilde{%
\Phi}S}_{x}{\bf \Phi })_{\alpha \alpha }\right| ={\bf \ \min }\sum_{\alpha
=1}^{N}\left| ({\bf \tilde{R}\tilde{W}S}_{x}{\bf WR})_{\alpha \alpha
}\right| ,
\end{equation}
minimized over all possible {\bf R }satisfying Eq. (26). Correspondingly
Eqs. (29) - (34) are pertinent to Eq. (78) provided: the subscript {\it x}
now is attached to the quantities {\it U}$_{W},\Delta _{W},\lambda $ and 
{\bf Q}; {\bf Q}$_{x}$ now is a 6$\times N$ matrix; the sums over {\it m} in
Eqs. (33) and (34) run from 1 to 6; and the six eigenvalues 
\begin{equation}
\lambda _{x1}\geq \lambda _{x2}\geq \lambda _{x3}\geq \lambda _{x4}\geq
\lambda _{x5}\geq \lambda _{x6}\geq 0
\end{equation}
of $\Delta _{Wx}$ are the square roots of the eigenvalues of $\rho ^{1/2}%
{\bf S}_{x}\rho ^{*}{\bf S}_{x}\rho ^{1/2}.$

The matrix ${\bf S}_{x}$ defined by Eq. (57) has two rows and two columns
that are identically zero. It readily can be seen that the corresponding
rows and columns of the matrix ${\bf S}_{x}\rho ^{*}{\bf S}_{x}$ are
identically zero as well. Thus the 6$\times 6$ Hermitian matrix ${\bf S}
_{x}\rho ^{*}{\bf S}_{x}$ has at least two zero eigenvalues, i.e., has [11]
a rank no greater than 4. It then follows [12], as can be verified by direct
multiplication, that the Hermitian matrix $\rho ^{1/2}{\bf S} _{x}\rho ^{*}%
{\bf S}_{x}\rho ^{1/2}$ also has rank no greater than 4, i.e., that at least 
$\lambda _{x5}$ and $\lambda _{x6}$ are zero in Eq. (79). Consequently,
except that subscripts {\it x} must be appropriately attached, Eq. (35)
remains valid as written; in particular Eq. (35) gives a lower bound on $%
{\bf C}_{x}{\bf (}\rho {\bf )}$ even though it does not include terms
proportional to $\lambda _{x5}$ and $\lambda _{x6}.$

Therefore, now replacing the two qubit 4$\times 4$ matrix of Eq. (37) with
the unitary 6$\times 6$ matrix 
\begin{equation}
{\bf Q}_{3>}=\frac{1}{2}\left( 
\begin{array}{cccccc}
1 & 1 & 1 & 1 & 0 & 0 \\ 
i & i & -i & -i & 0 & 0 \\ 
i & -i & i & -i & 0 & 0 \\ 
i & -i & -i & i & 0 & 0 \\ 
0 & 0 & 0 & 0 & 2 & 0 \\ 
0 & 0 & 0 & 0 & 0 & 2
\end{array}
\right) ,
\end{equation}
and obtaining the 6$\times 6$ ${\bf Q}_{3<}$ via the corresponding
replacement in Eq. (42), we conclude that 
\begin{equation}
{\bf C}_{x}{\bf (}\rho )={\bf \max (}0{\bf ,}\lambda _{x1}-\lambda
_{x2}-\lambda _{x3}-\lambda _{x4})
\end{equation}
is the the qubit-qutrit generalization of Eq. (43) for the quantity defined
by Eq. (74). Noting that each of the matrices ${\bf S}_{y}$ and ${\bf S}_{z}$
defined by Eqs. (58) and (59) also has two rows and two columns that are
identically zero, we evidently can further conclude that the quantities
defined by the corresponding Eqs. (75) and (76) are given by 
\begin{equation}
{\bf C}_{y}{\bf (}\rho )={\bf \max (}0{\bf ,}\lambda _{y1}-\lambda
_{y2}-\lambda _{y3}-\lambda _{y4}),
\end{equation}
\begin{equation}
{\bf C}_{z}{\bf (}\rho )={\bf \max (}0{\bf ,}\lambda _{z1}-\lambda
_{z2}-\lambda _{z3}-\lambda _{z4}),
\end{equation}
wherein the $\lambda _{y}$ and $\lambda _{z},$ each ordered as in Eq. (36),
are the square roots of the four largest eigenvalues (some of which may be
zero) of the respective matrices $\rho ^{1/2}{\bf S}_{y}\rho ^{*}{\bf S}%
_{y}\rho ^{1/2}$ and $\rho ^{1/2}{\bf S}_{z}\rho ^{*}{\bf S}_{z}\rho ^{1/2}.$

Define 
\begin{equation}
{\bf C}_{3b}{\bf (}\rho )=({\bf C}_{x}^{2}+{\bf C}_{y}^{2}+{\bf C}%
_{z}^{2})^{1/2},
\end{equation}
where the quantities on the right side of Eq. (84) have the values given by
Eqs. (81) - (83). Then ${\bf C}_{3b}{\bf (}\rho )$ is a lower bound on the
qubit-qutrit ${\bf C(}\rho )$ defined by Eq. (61). Correspondingly,
remembering Eq. (14), the desired lower bound on the qubit-qutrit EOF is $%
\varepsilon [{\bf C}_{3b}{\bf (}\rho )].$ It can be seen, much as discussed
for the two qubit case at the end of Section II.A. above, that the lower
bounds given by Eqs. (81)-(83): (i) are yielded by large ranges of 6$\times
N $ matrices {\bf Q}$_{x}${\bf ,Q}$_{y},${\bf Q}$_{z},${\bf \ }with {\it N}
arbitrarily large, and (ii) remain valid even when the qubit-qutrit $\rho $
has a rank {\it r }$\geq ${\it \ }2 but less than the value 6 implicitly
assumed in the foregoing derivations of those equations.

\section{The Qubit-Qudit System.}

We now turn to the derivation of the lower bound on the qubit-qudit EOF.
This derivation is a straightforward extension of the qubit-qutrit
derivation; there is no need to give as many details as were furnished
heretofore. The analog of Eq. (45) is 
\begin{equation}
A_{\alpha }=\left( 
\begin{array}{cccccc}
a_{11}^{\alpha } & a_{12}^{\alpha } & a_{13}^{\alpha } & ... & 
a_{1(d-1)}^{\alpha } & a_{1d}^{\alpha } \\ 
a_{21}^{\alpha } & a_{22}^{\alpha } & a_{23}^{\alpha } & ... & 
a_{2(d-1)}^{\alpha } & a_{2d}^{\alpha } \\ 
0 & 0 & 0 & ... & 0 & 0 \\ 
0 & 0 & 0 & ... & 0 & 0 \\ 
... & ... & ... & ... & ... & ... \\ 
0 & 0 & 0 & 0 & 0 & 0
\end{array}
\right) ,
\end{equation}
wherein the {\it d}$\times d$ matrix $A_{\alpha }$ is made square by
incorporating {\it d}-2 rows of zeros. The analog of Eq. (47) then is 
\begin{equation}
A_{\alpha }A_{\alpha }^{\dagger }=\left( 
\begin{array}{cccc}
b_{11}^{\alpha } & b_{12}^{\alpha } & ... & 0 \\ 
b_{21}^{\alpha } & b_{22}^{\alpha } & ... & 0 \\ 
... & ... & ... & ... \\ 
0 & 0 & 0 & 0
\end{array}
\right) ,
\end{equation}
wherein the only non-vanishing elements are $b_{11}^{\alpha },b_{12}^{\alpha
},b_{21}^{\alpha }$ and $b_{22}^{\alpha },$ given by: 
\begin{equation}
b_{11}^{\alpha }=\sum_{j=1}^{d}\left| a_{1j}^{\alpha }\right| ^{2}
\end{equation}
\begin{equation}
b_{12}^{\alpha }=\sum_{j=1}^{d}a_{1j}^{\alpha }a_{2j}^{\alpha *}
\end{equation}
\begin{equation}
b_{21}^{\alpha }=b_{12}^{\alpha *}
\end{equation}
\begin{equation}
b_{22}^{\alpha }=\sum_{j=1}^{d}\left| a_{2j}^{\alpha }\right| ^{2}.
\end{equation}
Thus $A_{\alpha }A_{\alpha }^{\dagger }$ has at most two non-vanishing
eigenvalues $(c_{1}^{\alpha })^{2},(c_{2}^{\alpha })^{2}$ which continue to
obey Eqs. (5) and (55). Eqs. (87) -(90) now yield 
\begin{equation}
(c_{1}^{\alpha })^{2}(c_{2}^{\alpha })^{2}=b_{11}^{\alpha }b_{22}^{\alpha
}-b_{12}^{\alpha }b_{21}^{\alpha }=\frac{1}{2}\sum_{i=1}^{d}\sum_{j=1}^{d}%
\left| a_{1i}^{\alpha }a_{2j}^{\alpha }-a_{1j}^{\alpha }a_{2i}^{\alpha
}\right| ^{2}=\sum_{j>i}^{d}\sum_{i=1}^{d}\left| a_{1i}^{\alpha
}a_{2j}^{\alpha }-a_{1j}^{\alpha }a_{2i}^{\alpha }\right| ^{2}.
\end{equation}
That Eqs. (87)-(90) imply Eq. (91) can be demonstrated by straightforward
algebraic manipulations, as well as via mathematical induction starting with
the presumed correctness of Eq. (91) when the sums over {\it i }and{\it \ j }
in Eqs. (87) - (91) run from 1 to {\it d}-1 only.

Let $\Psi _{\alpha }$ be the column matrix whose components $\Psi _{m\alpha
},$ {\it m = }1 to 2{\it d}, are the coefficients {\it a}$_{ij}^{\alpha },$
in the order $a_{11}^{\alpha },a_{12}^{\alpha },...,a_{1d}^{\alpha
},a_{21}^{\alpha },a_{22}^{\alpha },...,a_{2d}^{\alpha }.$ Define the set of 
{\it d(d-}1)/2 symmetric 2{\it d}$\times $2{\it d} square matrices {\bf S}$%
^{ij},1\leq i\leq d-1,$ {\it j} 
\mbox{$>$}%
{\it i}, to be the matrices whose elements {\bf S}$_{mn}^{ij}$ all are zero
except for: 
\begin{equation}
{\bf S}_{i,j+d}^{ij}={\bf S}_{j+d,i}^{ij}=1,
\end{equation}
\begin{equation}
{\bf S}_{j,i+d}^{ij}={\bf S}_{i+d,j}^{ij}=-1.
\end{equation}
Then 
\begin{equation}
\tilde{\Psi}_{\alpha }{\bf S}^{ij}\Psi _{\alpha }=2(a_{1i}^{\alpha
}a_{2j}^{\alpha }-a_{1j}^{\alpha }a_{2i}^{\alpha }),
\end{equation}
and 
\begin{equation}
{\bf C(}\rho {\bf )=\min }\sum_{\alpha =1}^{N}p_{\alpha }Con(\Psi _{\alpha
})={\bf \min }\sum_{\alpha =1}^{N}\left( \sum_{j>i}\sum_{i=1}^{d-1}\left| 
\tilde{\Phi}_{\alpha }{\bf S}^{ij}\Phi _{\alpha }\right| ^{2}\right) ^{1/2}.
\end{equation}
Further defining 
\begin{equation}
{\bf C}_{ij}{\bf (}\rho {\bf )=\min }\sum_{\alpha =1}^{N}\left| \tilde{\Phi}%
_{\alpha }{\bf S}^{ij}\Phi _{\alpha }\right| ,
\end{equation}
Eq. (77) generalizes to 
\begin{equation}
{\Large \lbrack }\sum_{j>i}\sum_{i=1}^{d-1}{\bf C}_{ij}^{2}{\bf (}\rho {\bf )%
}{\Large ]}^{\frac{1}{2}}\leq {\bf C(}\rho {\bf ).}
\end{equation}

The left side of Eq. (97) is the desired lower bound ${\bf C}_{db}(\rho )$
on the qubit-qudit ${\bf C(}\rho {\bf )}$, using 
\begin{equation}
{\bf C}_{ij}{\bf (}\rho {\bf )}={\bf \max (}0{\bf ,}\lambda
_{1}^{ij}-\lambda _{2}^{ij}-\lambda _{3}^{ij}-\lambda _{4}^{ij}),
\end{equation}
wherein the $\lambda ^{ij},$ ordered as in Eq. (36), are the square roots of
the four largest eigenvalues of the matrix $\rho ^{1/2}{\bf S}^{ij}\rho ^{*}%
{\bf S}^{ij}\rho ^{1/2}.$ The desired lower bound on the qubit-qudit EOF is $%
\varepsilon [{\bf C}_{db}{\bf (}\rho )].$

\section{Concluding Remarks.}

It is needful first to point out that in a number of respects our review of
Wootters' two qubit derivation [2] has echoed the treatment given by
Audenaert {\it et al }[13]; in particular these authors have made use of the
matrix notation for decompositions of $\rho $ introduced in Eqs. (24) and
(25), and have noted that the minimization over all {\bf R} in Eq. (28) can
be replaced by the more convenient minimization over all {\bf Q} in Eq.
(33). Also the matrices {\bf S}$^{ij}$ appearing in Eqs. (92)-(96)
essentially are the ''indicator matrices'' introduced by Audenaert {\it et
al }[13], as those matrices would be written for the qubit-qutrit system.
For the purpose of generalizing Wootters' derivation to obtain the
qubit-qutrit EOF lower bound, however, we have found it preferable not to
rely on Thompson's Theorem [14] as Audenaert {\it et al }[13] do, but rather
to explicitly construct the two qubit matrices ${\bf Q}_{>}$ and ${\bf Q}%
_{<} $ of Eqs. (37) and (42). This procedure enables our relatively
straightforward derivations of the lower bound Eqs. (84) and (98), on the
average qubit-qutrit and qubit-qudit concurrences respectively. Our
procedure also makes apparent the fact that (as discussed at the end of
Section II.B.) for any qubit-qutrit $\rho $ there always actually exists a
wide range of matrices {\bf Q}$_{x},$ i.e., a wide range of decompositions
of $\rho ,$ having an average ''{\it x-}concurrence'' $\sum_{\alpha }\left| 
\tilde{\Phi}_{\alpha }{\bf S}_{x}\Phi _{\alpha }\right| $ which actually
attains the minimum possible value ${\bf C}_{x}{\bf (}\rho {\bf )}$ given by
Eq. (81), and similarly for ${\bf C} _{y}{\bf (}\rho {\bf )}$ and ${\bf C}%
_{z}{\bf (}\rho {\bf ).}$

We have performed no numerical calculations designed to estimate the utility
of our lower bounds. Nor have we attempted, by numerical calculation or
otherwise, to ascertain whether these just described three different ranges
of decompositions of $\rho ,$ which separately attain ${\bf C}_{x}(\rho )%
{\bf ,}$ ${\bf C}_{y}{\bf (}\rho {\bf )}$ and ${\bf C}_{z}{\bf (}\rho {\bf )}
$ respectively, generally will overlap sufficiently to ensure the
attainability of the bound given by Eq. (84). In other words we have not
tried to answer the question: Given some arbitrary qubit-qutrtit $\rho ,$
will it generally be possible to find a decomposition of $\rho $ whose
minimum average concurrence equals the value of ${\bf C}_{3b}{\bf (}\rho 
{\bf )}$ given by Eq. (84)? If such a decomposition exists, then [recalling
the two qubit definition of {\bf Q }given immediately preceding Eq. (32)] it
must be possible to find three matrices {\bf Q}$_{x},${\bf Q}$_{y},${\bf Q}$%
_{z}$ such that 
\begin{equation}
U_{Wx}^{\dagger }{\bf Q}_{x}=U_{Wy}^{\dagger }{\bf Q}_{y}=U_{Wz}^{\dagger }%
{\bf Q}_{z}={\bf R}_{b},
\end{equation}
where: {\bf Q}$_{x},${\bf Q}$_{y},${\bf Q}$_{z}$ are unitary 6$\times N$
matrices (all with the same $N)$ which, when employed in the qubit-qutrit
analogs of Eqs. (33)-(35), imply Eqs. (81)-(83) respectively; $%
U_{Wx},U_{Wy},U_{Wz}$ are the properly specified analogs of the Eq. (29) $%
U_{W}$ when {\bf S} in Eq. (29) is replaced by {\bf S}$_{x}${\bf ,S}$_{y}$%
{\bf ,S}$_{z}$ respectively (because $\lambda _{x5}=\lambda _{x6}=0$ in Eq.
(79), $U_{Wx}$ is not uniquely determined by the 6$\times 6$ ${\bf S}_{x}$
analog of Eq. (29), and similarly for $U_{Wy}$ and $U_{Wz})$; and the 6$%
\times N{\bf \ }$matrix ${\bf R}_{b}$ is the ${\bf R}$ satisfying Eq. (26)
which, via the qubit-qutrit analog of Eq. (27), specifies this particular
decomposition.

Evidently it must be possible to find a set of matrices {\bf Q}$_{x},${\bf Q}%
$_{y},${\bf Q}$_{z}$ satisfying Eq. (99) whenever the given qubit-qutrit
matrix $\rho $ happens to be separable, i.e., whenever {\bf E}${\bf (}\rho
)=0,$ in which event both ${\bf C(}\rho )$ and ${\bf C}_{3b}{\bf (}\rho )$
also must vanish. On the other hand, numerical calculations by Audenaert 
{\it et al }[13] suggest that it may not be possible to find such {\bf Q}$%
_{x},${\bf Q}$_{y},${\bf Q}$_{z}$ for every qubit-qutrit $\rho $ with ${\bf C%
}_{3b}{\bf (}\rho )=0.$ In short, whereas each of ${\bf C}_{x}{\bf (}\rho 
{\bf )}=0,{\bf C}_{x}{\bf (}\rho {\bf )}=0{\bf ,}$ ${\bf C}_{x}{\bf (}\rho 
{\bf )}=0$ surely is a necessary condition for separability (as Audenaert 
{\it et al }[13] already had concluded, though via a quite different
formalism than employed herein), the totality of these three conditions may
not suffice to guarantee separability. Moreover this writer knows of no
theoretical or numerical studies which might answer the question posed in
the preceding paragraph under the circumstance that ${\bf C}_{3b}{\bf (}\rho
)>0.$ In his opinion, even the demonstrated inability to satisfy Eq. (99)
for most qubit-qutrit $\rho $ with ${\bf C}_{3b}{\bf (}\rho )=0$ would not
imply the inability to satisfy Eq. (99) for most $\rho $ with ${\bf C}_{3b}%
{\bf (}\rho )>0,$ especially when ${\bf C}_{3b}{\bf (}\rho )>0$ results from
the non-vanishing of each of ${\bf C}_{x}{\bf (}\rho {\bf ),}$ ${\bf C}_{y}%
{\bf (}\rho {\bf )}$ and ${\bf C}_{z}{\bf (}\rho {\bf ).}$ This opinion
stems from the observation that the range of matrices {\bf Q}$_{x}$ yielding
a specified value of ${\bf C}_{x}{\bf (}\rho {\bf )}$ 
\mbox{$>$}%
0 is wider than the range of {\bf Q}$_{x}$ yielding ${\bf C}_{x}{\bf (}\rho 
{\bf )}=0,$ reflecting the fact that achieving Eq. (41) requires $%
\sum_{m=1}^{4}\lambda _{m}{\bf Q}_{m\alpha }^{2}$ $=0$ for every one of the 
{\it N} $\alpha -$indexed columns of ${\bf Q},$ whereas Eq. (38) can be
achieved with values of $\sum_{m=1}^{4}\lambda _{m}{\bf Q}_{m\alpha }^{2}$
which are unequal for different $\alpha ,$ 1 $\leq \alpha \leq N.$

Our Eqs. (56) and (91), which are the crucial starting points for our
derivations of the lower bounds ${\bf C}_{3b}{\bf (}\rho )$ and ${\bf C}_{db}%
{\bf (}\rho )$ given by Eqs. (84) and (97) respectively, could have been
deduced more elegantly but less transparently from the equalities 
\begin{equation}
1-Tr(\rho _{A}^{2}{\bf )}={\bf [}Tr(\rho _{A}){\bf ]}^{2}-Tr(\rho _{A}^{2}%
{\bf )=}\text{ }\left( \sum_{k=1}^{K}c_{k}^{2}\right)
^{2}-\sum_{k=1}^{K}c_{k}^{4}=2\sum_{l>k}^{K}\sum_{k=1}^{K}c_{k}^{2}c_{l}^{2}%
{\bf =}  \eqnum{100a}
\end{equation}

\begin{equation}
{\bf =}\sum_{i,m=1}^{d_{A}}\sum_{j,n=1}^{d_{B}}\left(
a_{ij}a_{ij}^{*}a_{mn}a_{mn}^{*}-a_{ij}a_{mj}^{*}a_{mn}a_{in}^{*}\right)
=\sum_{i,m=1}^{d_{A}}\sum_{j,n=1}^{d_{B}}a_{ij}a_{mn}\left(
a_{ij}^{*}a_{mn}^{*}-a_{mj}^{*}a_{in}^{*}\right) =  \eqnum{100b}
\end{equation}

\begin{equation}
=\frac{1}{2}\sum_{i,m=1}^{d_{A}}\sum_{j,n=1}^{d_{B}}\left|
a_{ij}a_{mn}-a_{in}a_{mj}\right|
^{2}=2\sum_{m>i}^{d_{A}}\sum_{i=1}^{d_{A}}\sum_{n>j}^{d_{B}}%
\sum_{j=1}^{d_{B}}\left| a_{ij}a_{mn}-a_{in}a_{mj}\right| ^{2}.  \eqnum{100c}
\end{equation}
In Eqs. (100): $\rho $ = $\Psi _{\alpha }\Psi _{\alpha }^{\dagger }$
(elements $\rho _{ij,mn}=$ $a_{ij}a_{mn}^{*}),$ is a pure state density
matrix in a bipartite system composed of arbitrary subsystems A and B (not
merely a system composed of a qubit subsystem A and a qudit subsystem B);
the wave function $\Psi _{\alpha }$ has the expansion given by Eq. (2),
except that $i$ and $j$ now run from 1 to $d_{A}$ and $d_{B}$ respectively,
where $d_{A},d_{B}$ are the respective dimensionalities (i.e., the maximum
number of independent orthonormal eigenfunctions) in subsystems A, B; $\rho
_{A}$ (elements ($\rho _{A})_{im}=\sum_{j=1}^{d_{B}}a_{ij}a_{mj}^{*}),$ is
the reduced density matrix associated with subsystem A; the {\it K} real
quantities $c_{k}^{2}$ are the non-vanishing eigenvalues of $\rho _{A};$ the
here superfluous index $\alpha $ on the quantities $a_{ij}^{\alpha }$ and $%
c_{k}^{\alpha }$ of Eqs. (2) and (4) has been dropped; and the elements ($%
\rho _{A})_{im}$ for a qubit-qudit system can be seen to be identical with
the elements {\it b}$_{im}$ of Eqs. (48)-(51). For the qubit-qudit system,
where {\it K} = 2, the equality between the last term in Eq. (100a) and the
last term in Eq. (100c) is Eq. (91).

Eqs. (100), which in essence have been stated by Rungta {\it et al }[15] and
by Albeverio and Fei [16], are the basis for those authors' proposed
generalizations of Wootters' [2] two qubit concurrence. In particular,
except possibly for here inconsequential numerical factors, the left side of
Eq. (100a) is the square of the ''I-concurrence'' defined by Rungta {\it et
al }[15], while the right side of Eq. (100c) is the square of the
''generalized concurrence'' introduced by Albeverio and Fei [16]; the right
side of Eq. (100c) also is the ''length'' squared of the ''concurrence
vector'' introduced by Audenaert {\it et al }[13], again except for a here
inconsequential numerical factor. The numerical equality of these different
proposed generalizations of the two qubit concurrence, remarked on by
Wootters [1], is obvious from Eqs. (100). The present paper shows that these
generalizations are appropriate and useful in qubit-qudit systems, where the
von Neumann entropy of a pure state is a function of a single real parameter
only. On the other hand, the presence in Eq. (100a) of terms in $%
c_{k}^{2}c_{l}^{2}$ beyond $c_{1}^{2}c_{2}^{2}$ when {\it K} 
\mbox{$>$}%
2 strongly suggests that the utility of the aforementioned concurrence
generalizations for estimating the EOF in bipartite systems not containing a
qubit will be significantly less than in the {\it K} = 2 qubit-qudit systems
which are the subject of this paper.

Ultimately the analytical difficulties engendered by the square roots in Eq.
(61), remarked on in connection with that equation, stem from the decision
that the right sides of Eqs. (8) and (9) are to be regarded as functions of 
{\it C}$_{\alpha }$ rather than as functions of {\it C}$_{\alpha }^{2}, $
i.e., as functions of the concurrence rather than the square of the
concurrence. Focusing on {\it C}$_{\alpha }^{2}$ would have led to
replacement of Eq. (61) by 
\begin{equation}
\tau (\rho )={\bf \min }\sum_{\alpha =1}^{N}p_{\alpha }\left( \left| \tilde{%
\Psi}_{\alpha }{\bf S}_{x}\Psi _{\alpha }\right| ^{2}+\left| \tilde{\Psi}
_{\alpha }{\bf S}_{y}\Psi _{\alpha }\right| ^{2}+\left| \tilde{\Psi}_{\alpha
}{\bf S}_{z}\Psi _{\alpha }\right| ^{2}\right) ,  \eqnum{101}
\end{equation}
which avoids the troublesome square roots. The quantity $\tau (\rho ),$ here
denoting the minimum average square concurrence that can be attained by any
decomposition of $\rho ,$ is known as the ''tangle'' [17]. Unfortunately
this writer knows of no way to express the tangle of an arbitrary
qubit-qutrit $\rho $ in the form of readily computable properties of
matrices simply related to $\rho ,$ e.g., in terms of the eigenvalues of $%
\rho ^{1/2}{\bf S}_{x}\rho ^{*}{\bf S}_{x}\rho ^{1/2},$ as in Eqs. (81)-(83)
yielding the lower bound Eq. (84). Osborne [17] has found a closed form
expression for the tangle of any bipartite density matrix of rank {\it r = }
2, and [noting that the von Neumann entropy of Eq. (6) is a monotonically
increasing concave (not convex) function of {\it C}$_{\alpha }^{2}$ via Eqs.
(8)-(9)], has used this expression to deduce a very close upper bound for
the EOF of a rank 2 qubit-qudit $\rho .$ We have put no rank limitations on
our qubit-qudit density matrices.

Finally, it hardly is necessary to note that any of the foregoing remarks
which were made solely in a qubit-qutrit context have obvious
generalizations to the qubit-qudit system.

\smallskip

\_\_\_\_\_\_\_\_\_\_\_\_\_\_\_\_\_\_\_\_\_\_\_\_\_\_\_\_\_\_\_\_\_\_\_\_\_\_%
\_\_\_\_\_\_\_\_\_\_\_\_

{\footnotesize [1] W. K. Wootters, Quant. Inf. and Comp. {\bf 1}, 27 (2001).}

{\footnotesize [2] W. K. Wootters, Phys. Rev. Lett. {\bf 80}, 2245 (1998).}

{\footnotesize [3] L. P. Hughston, R. Josza, and W. K. Wootters, Phys. Lett.
A {\bf 183}, 14 (1993).}

{\footnotesize [4] M. A. Nielsen and I. L. Chuang, {\it Quantum Computation
and Quantum Information} (Cambridge University Press, Cambridge, England,
2000), p. 109.}

{\footnotesize [5] Nielsen and Chuang, {\it ibid}, p. 79.}

{\footnotesize [6] R. A. Horn and C. R. Johnson,{\it \ Matrix Analysis}
(Cambridge University Press, New York, 1985), pp. 414 ff.}

{\footnotesize [7] Horn and C. R. Johnson, {\it ibid}, p.{\it \ 7.}}

{\footnotesize [8] Horn and C. R. Johnson, {\it ibid}, pp. 40-42.}

{\footnotesize [9] Horn and C. R. Johnson, {\it ibid}, pp. 204-205.}

{\footnotesize [10] Horn and C. R. Johnson, {\it ibid}, pp 396 and 402.}

{\footnotesize [11] Horn and C. R. Johnson, {\it ibid}, p. 175, problem 12.}

{\footnotesize [12] Horn and C. R. Johnson, {\it ibid}, p. 13.}

{\footnotesize [13] K. Audenaert, F. Verstraete, and B. De Moor,
quant-ph/0006128 (2000).}

{\footnotesize [14] R. C. Thompson, {\it Linear Algebra and its Applications}
{\bf 26}, 65-106 (1979).}

{\footnotesize [15] P. Rungta, V. Buzek, C. M. Caves, M. Hillery, and G. J.
Milburn, Phys. Rev. A {\bf 64}, 043215 (2001).}

{\footnotesize [16] S. Albeverio and S. M. Fei, quant-ph/0109073 (2001).}

{\footnotesize [17] T. J. Osborne, quant-ph/0203087 (2002). }

\end{document}